\documentclass[aps,prb,twocolumn,floatfix,amsmath,amssymb]{revtex4}
\usepackage[final]{graphicx}
\usepackage{bm}
\usepackage{float}
\usepackage{afterpage}

\begin{document}

\title{Dimerization, Trimerization and Quantum pumping}

\author{Huaiming Guo}

\affiliation{Department of Physics, Beihang University, Beijing, 100191, China}

\begin{abstract}
We study one-dimensional topological models with dimerization and trimerization and show that these models can be generated using interaction or optical superlattice. The topological properties of these models are demonstrated by the appearance of edge states and the mechanism of dimerization and trimerization is analyzed. Then we show that a quantum pumping process can be constructed based on each one-dimensional topological model. The quantum pumping process is explicitly demonstrated by the instantaneous energy spectrum and local current. The result shows that the pumping is assisted by the gapless states connecting the bands and one charge is pumped during a cycle, which also defines a nonzero Chern number. Our study systematically shows the connection of one-dimensional topological models and quantum pumping, and is useful for the experimental studies on topological phases in optical lattices and photonic quasicrystals.
\end{abstract}


\maketitle

\section{Introduction}
The discovery of topological insulators have aroused the interests in the study of topological phase of matter \cite{rev1,rev2,rev3,rev4}. Specifically recently a lot of works focus on one dimension (1D) \cite{rev5,rev6,rev7,rev8,rev9,rev10,rev11,rev12,rev13,rev14,rev15,rev16}. On the one hand the physics of 1D is as rich as higher dimensions, but easier to understand. On the other hand many 1D systems can be realized experimentally in ultracold atomic system \cite{rev13,rev14,rev15} and photonic quasicrystals \cite{rev12}, thus provide platforms to test the theoretical predictions.

 In 1979, Su et al. suggest a model to study the solitons in polyacetylene \cite{rev17}, which has become a famous 1D topological model and many studies have been carried on it. Later Su and Schrieffer also study a 1D one-third filled Peierls system and find the fractional topological excitation associated with the kinks \cite{rev18}. The above two models are also known as dimerized and trimerized models. The physical properties of the two models represent   a new kind of order called topological order \cite{rev19}, which is beyond Landau symmetry breaking theory. Generally the topological order contains two different types: the intrinsic and symmetry protected ones. In 1D, only symmetry protected topological order exists.

Recently there appear works studying the topological phases in 1D optical superlattices, including topological phase transition, fractional topological phases, topological Mott insulators, and even topological superconductor \cite{rev5,rev8,rev11}. The based model of the above studies is simple, but the phenomena it exhbits are rich. More inspiringly, this model has been realized using photonic quasicrystals \cite{rev12} and is very possible to be realized in the optical lattices. These theoretical and experimental achievements motivate us to have more understanding on this 1D model and related physics.

In this paper, we show that the topological phase induced by optical superlattice is closely related to the trimerized model, and besides the interaction can generate topological phase as that in the dimerized model. We also connect the 1D topological phase with the quantum pumping and the Chern number. These results provide more understanding on the 1D topological phase and  are useful for the experimental studies on topological phases in optical lattices and photonic quasicrystals. The paper is organized as follows. In Sec.2, the dimerized and trimerized models are revisited and the topological properties are demonstrated with edge states on open chains. In Sec.3, we generate models exhibiting similar properties as the dimerized and trimerized models using the interaction and optical superlattice and discuss the reason for that. In Sec.4, We show that based on each 1D topological model, a quantum pumping process can be constructed, which is explicit shown by the instantaneous energy spectrum and local current, and in the process a nonzero Chern number is defined. Finally we conclude the work in Sec.5.
\section{The models}
The first model is with the dimerization, which writes \cite{rev17},
\begin{eqnarray}\label{eq1}
H_1=\sum_{i}(t_0+\delta t)c_{Ai}^{\dagger}c_{Bi}+(t_0-\delta t)c_{Ai+1}^{\dagger}c_{Bi}+h.c.,
\end{eqnarray}
where $c_{i}^{\dagger} (c_{i})$ is the creation (annihilation) operator of the fermion, and $A, B$ represent two different sites of a unit-cell. The hopping amplitude $t_0$ is set to be the unit of the energy ($t_0=1$). In the reciprocal space, the Hamiltonian is written as $H_1=\sum_{k}\psi_{k}^{\dagger}H_1(k)\psi_{k}$ with $\psi_{k}=(c_{A k},c_{Bk})^{T}$ and
\begin{eqnarray*}
H_1(k)=2t_0\cos{k}\sigma_{x}-2\delta t\sin{k}\sigma_{y},
\end{eqnarray*}
with $\sigma_{x,y}$ the Pauli matrices. The energy spectrum is given by
\begin{eqnarray*}
  E_{\bf k} &=& \pm \sqrt{4t_0^{2}\cos^2 k+4\delta t^2\sin^2 k},
\end{eqnarray*}
with the gap $4|\delta t|$ at the Dirac point $k=\pi/2$ (or $3\pi/2$). There are two topologically distinct phases depending on the sign of $\delta t$: for $\delta t>0$ it is topological trivial; for $\delta t<0$ the system shows topological property with zero-energy states localized near the ends. The topological invariant characterizing the two phases is Berry phase which is zero for $\delta t>0$ and $\pi$ for $\delta t<0$.

The second model we consider is with the trimerization \cite{rev18},
\begin{eqnarray}\label{eq2}
H_2=\sum_{i}t_1c_{Ai}^{\dagger}c_{Bi}+t_2c_{Bi}^{\dagger}c_{Ci}+t_3c_{Ci}^{\dagger}c_{Ai+1}+h.c. .
\end{eqnarray}
In this model, each unit-cell consists of three sites, $A, B$ and $C$. There are three configurations with perfect trimerization: 1, $t_1=t_2=t_0-\delta t$ and $t_3=t_0+\delta t$; 2, $t_2=t_3=t_0-\delta t$ and $t_1=t_0+\delta t$; 3, $t_1=t_3=t_0-\delta t$ and $t_2=t_0+\delta t$. In the following we discuss the first case and the others are similar. In the momentum space, the Hamiltonian writes as $H_2=\sum_{k}\psi_{k}^{\dagger}H_2(k)\psi_{k}$ with $\psi_{k}=(c_{A k},c_{Bk},c_{Ck})^{T}$ and
\begin{eqnarray*}
H_2(k)=(\begin{array}{ccc}
               0 & t_0-\delta t & (t_0+\delta t)e^{-ik} \\
               t_0-\delta t & 0 & t_0-\delta t \\
               (t_0+\delta t)e^{ik} & t_0-\delta t & 0
             \end{array})
\end{eqnarray*}
For $\delta t\neq 0$ the system is gapped at $\frac{1}{3}$ and $\frac{2}{3}$ filling and both the gaps are $\frac{3(t_0+\delta t)}{2}-\frac{1}{2}\sqrt{9t_0^2-14t\delta t+9\delta t^2}$. For $\frac{1}{3}$ filling it is located at $k=0$ while at $k=\pi$ for $\frac{2}{3}$ filling. Depending on the sign of $\delta t$ and the value of the chain length $N mod 3$, there appear different edge states. When $mod(N,3)=0$, there appear edge states at both ends for $\delta t>0$ but none for $\delta t<0$. When $mod(N,3)=1$, there appear one edge state at the left end for $\delta t>0$ but none for $\delta t<0$. When $mod(N,3)=2$, there appear one edge state at the left end for $\delta t>0$ but at the right end for $\delta t<0$.

\begin{figure}[htbp]
\centering
\includegraphics[width=8.5cm]{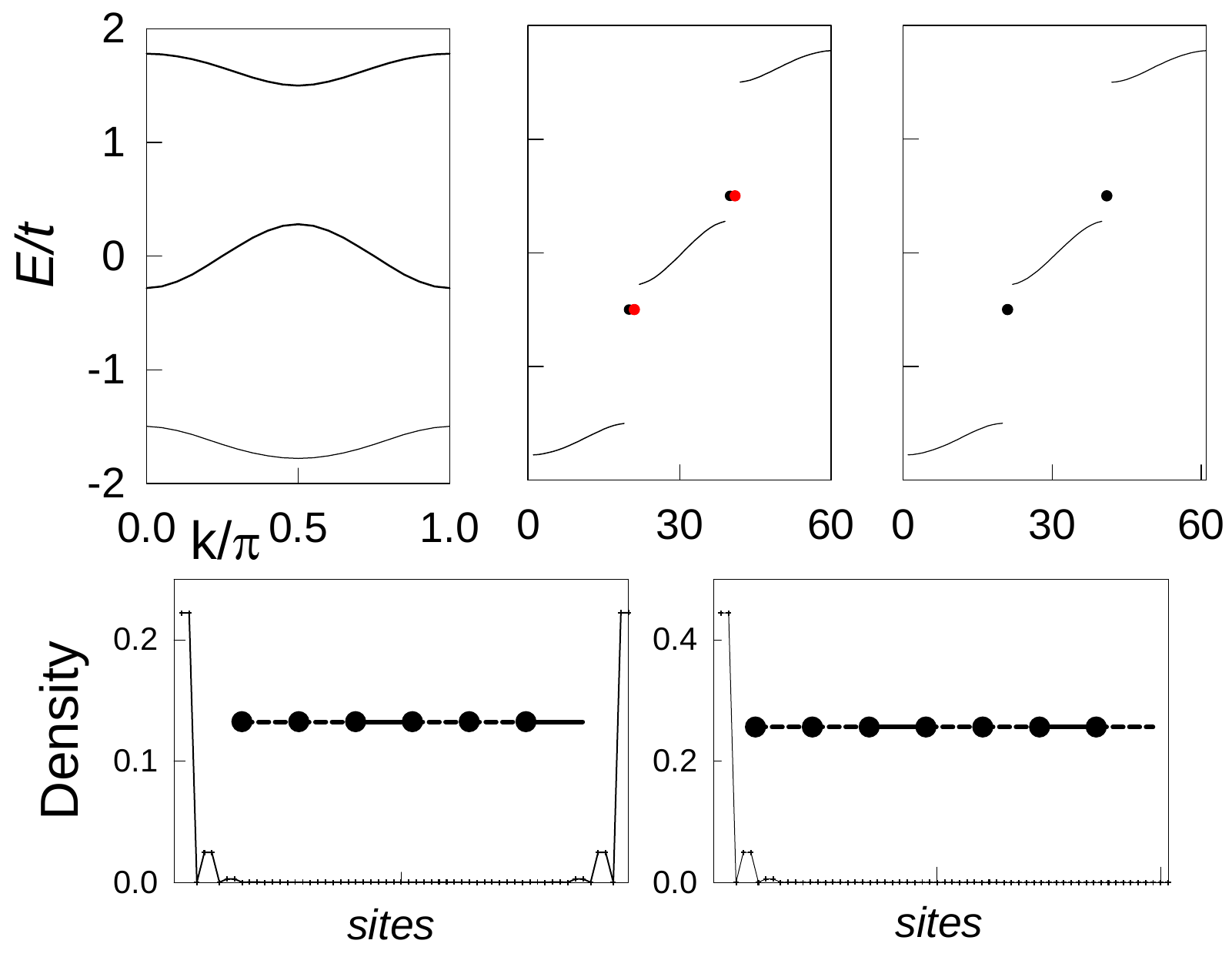}
\caption{(Color online)The energy spectrum of Hamiltonian Eq.(\ref{eq2}) with $\delta t=0.5$: (a) periodic boundary condition with $N=3N_u$; (b) and (c) open boundary condition with $N=3N_u$ and $N=3N_u+1$ respectively. Here $N_u=20$ is the number of unit-cells. (d) and (e): the distribution of the lower in-gap states in (b), (c) respectively. }\label{trimerize}
\end{figure}

\section{Generating $\delta t$ from interaction or optical superlattice}
In the previous section, we revisit the dimerized and trimerized models. Next we drop $\delta t$ terms in both models and dynamically generate $\delta t$ from interaction or optical superlattice.

\begin{figure}[htbp]
\centering
\includegraphics[width=8.5cm]{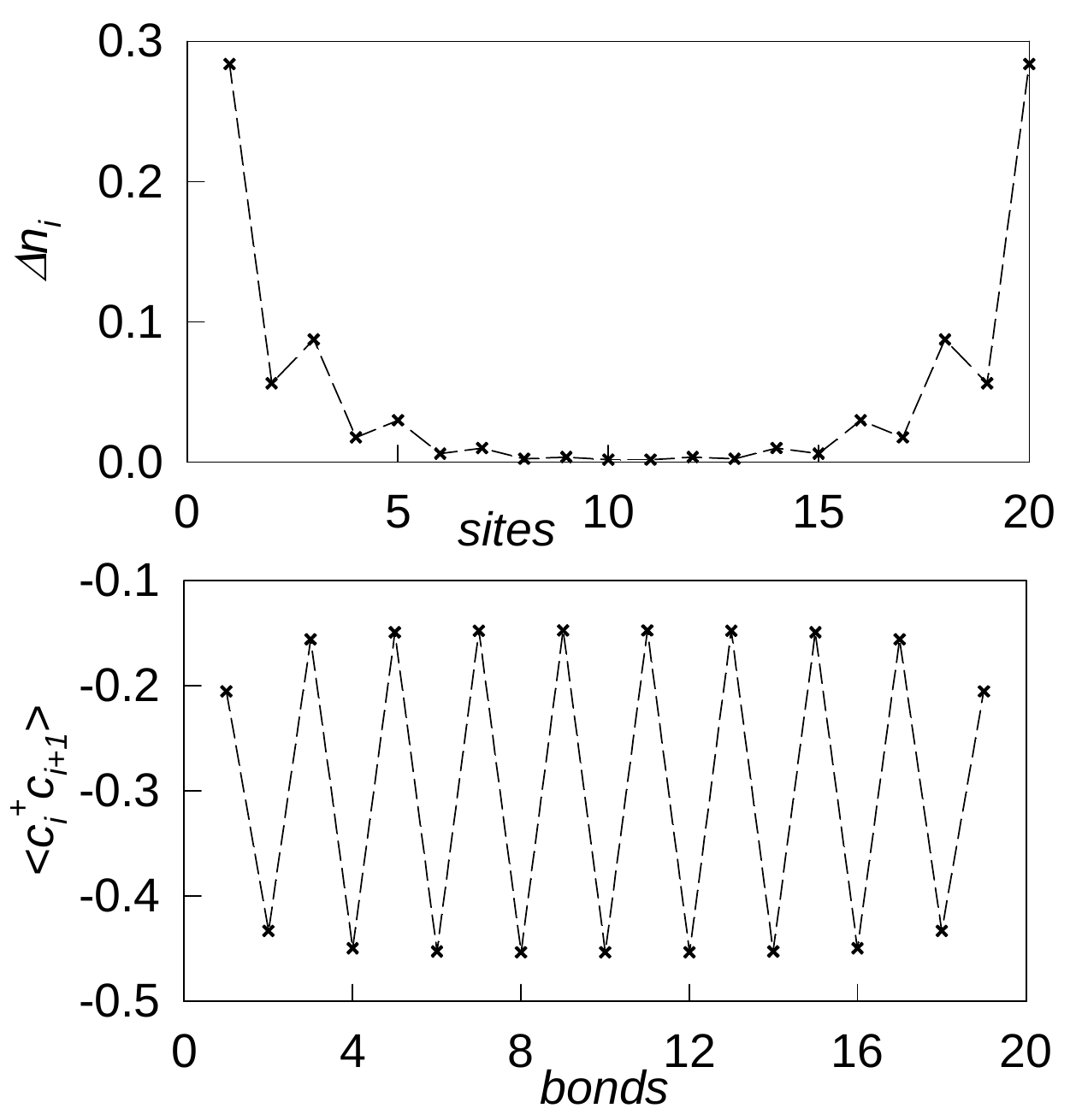}
\caption{(a) The distribution of the quasiparticles added
or removed from the half-filling system with OBC. (b) the average value of the hopping amplitude of each bond on an open chain. The calculations are for model Eq.(\ref{eq3}) with $V_1=1$ and the length of the chain $N=20$. }\label{fig2}
\end{figure}

We firstly generate a model showing similar properties to Eq.(\ref{eq1}) from nearest neighbor (NN) interaction. The Hamiltonian under consideration is:
\begin{eqnarray}\label{eq3}
H'_1=t_0\sum_{i}(c_i^{\dagger}c_{i+1}+h.c.)+\sum_{\langle ij\rangle}V_{ij} n_i n_j,
\end{eqnarray}
where the NN interaction is with alternating strength, i.e., $V_{ij}=(-1)^i V_{1}$ supposing $i<j$. For $V_1<0$ the system is trivial, while for $V_1>0$ the system becomes nontrivial when the system shows similar properties as Eq.(\ref{eq1}) at $\delta t<0$. For a many-body system, the topological properties can manifest itself from the edge states of quasiparticles. The energy of the quasiparticle added to a system with $n$ electrons can be defined as
$\Delta E_n =E_{n+1}^0-E_n^0$, where $E_n^0$ is the ground energy of a system with $n$ particles \cite{rev7}. For the case of $V_1>0$ there appear states in the gap of the quasiparticle energy spectrum as the boundary condition changes from periodic (PBC) to open (OBC) ones. The distribution of the in-gap state can be calculated which is defined as: $\Delta n_{i}=\langle
\psi^{0}_{n+1}|\hat{n}_{i}|\psi^{0}_{n+1}\rangle-\langle \psi^{0}_{n}|\hat{n}%
_{i}|\psi^{0}_{n}\rangle$, where $\hat{n}_{i}=c^{\dagger}_{i}c_{i}$ is the electron number
operator on site $i$ and $\psi^{0}_{n}$ is the ground-state wave function of the
system with $n$ electrons \cite{rev7}. The distribution of the in-gap state is shown in Fig.\ref{fig2}, which mainly distributes near the edges. So for the case of $V_1>0$ the system exhibits nontrivial topological properties.

We further calculate the average value of the hopping amplitude of each bond, which provides an explanation of the nontrivial phase in Eq.(\ref{eq3}). As shown in Fig.\ref{fig2}, due to the existence of the alternating NN interactions, the effective hopping amplitude $\langle c^{\dagger}_{i}c_{i+1}\rangle$ becomes alternating, which let Eq.(\ref{eq3}) show similar properties as the dimerized model Eq.(\ref{eq1}).

\begin{figure}[htbp]
\centering
\includegraphics[width=8.5cm]{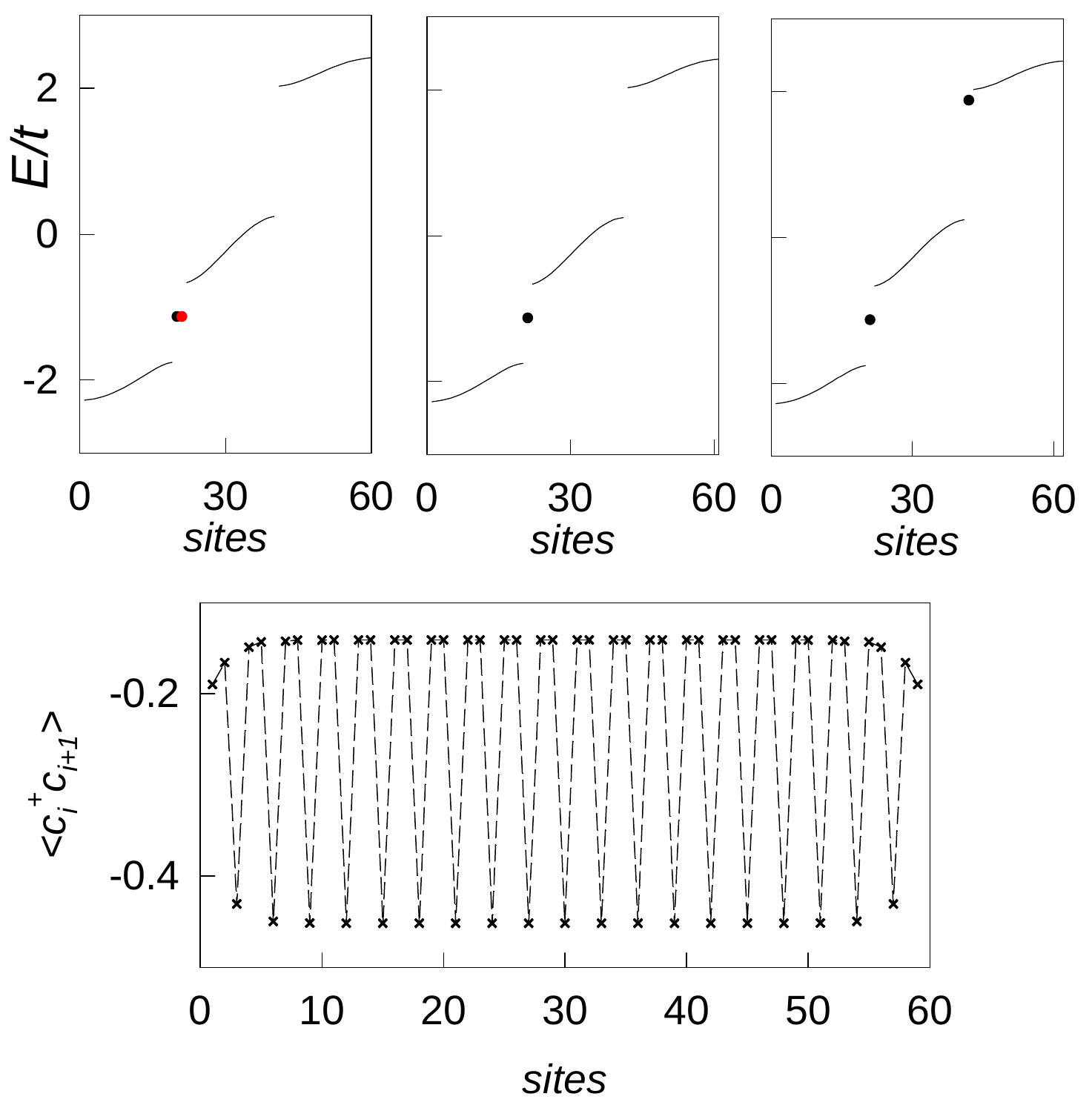}
\caption{(Color online) The energy spectrum of Hamiltonian Eq.(\ref{eq4}) with $V_2=1.5$ on an open chain with the lengths: (a) $N=3N_u$; (b) $N=3N_u+1$; (c) $N=3N_u+2$. (d) the average value of the hopping amplitude of each bond on an open chain with the length $N=3N_u$. Here $N_u=30$ is the number of unit-cells.}\label{fig3}
\end{figure}

Next we effectively generate Hamitonian Eq.(\ref{eq2}) from a model with a uniform nearest neighbor hopping and an optical superlattice, which writes,
\begin{eqnarray}\label{eq4}
H'_2=t_0\sum_{i}(c_i^{\dagger}c_{i+1}+h.c.)+\sum_{i=1}^{N}V_i n_i,
\end{eqnarray}
with $V_i=V_2\cos(\frac{2\pi}{3} i+\frac{2\pi}{3})$. With this superlattice a unit-cell still consistent of three sites $A, B, C$, and the on-site potentials are $V_A=-V_2/2, V_B=V_2, V_C=-V_2/2$ respectively. We calculate the energy spectrum of Eq.(\ref{eq4}) on an open chain with the lengths $N=3N_u, N=3N_u+1, N=3N_u+2$ where $N_u$ is the number of the unit-cells. As shown in Fig.\ref{fig3} (a), (b), (c), the results are similar to those of Hamiltonian Eq.(\ref{eq2}) at low-filling. The reason is that the potentials on sites $A, C$ are equal and lower, the particles tend to reside on these sites. So at low-filling the particle has equal probability residing on $A, C$, which makes the hopping amplitude between $A$ and $C$ stronger. We calculate the average value of the NN hopping amplitude $\langle c_i^{\dagger}c_{i+1}\rangle$. The average hopping amplitudes show a configuration with trimerization, which is consistent with the above qualitative analysis. Thus Eq.(\ref{eq2}) is dynamically generated using an optical superlattice.

\section{Quantum pumping in the models}

\begin{figure}[htbp]
\centering
\includegraphics[width=8.5cm]{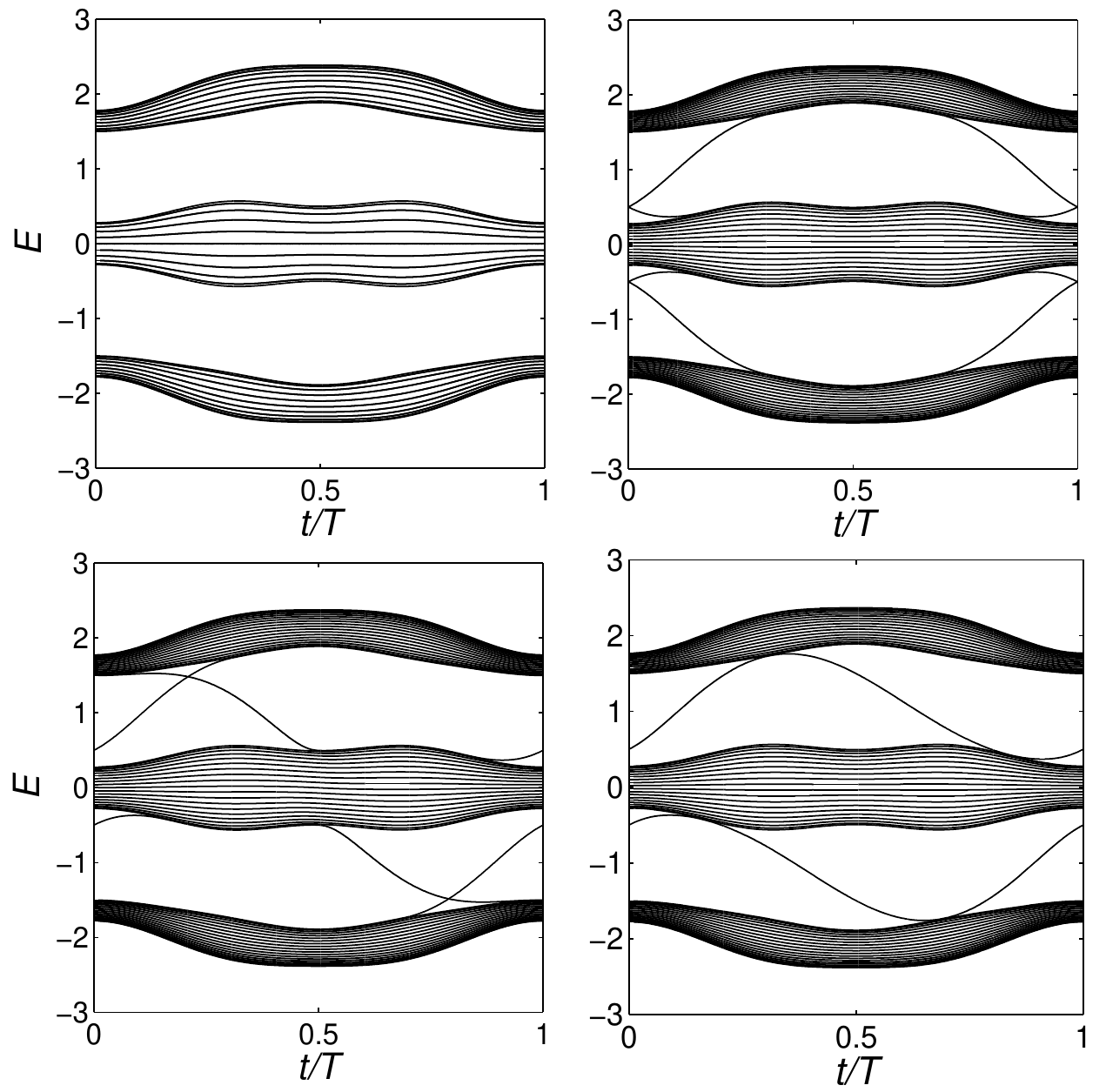}
\caption{(a) The instantaneous energy spectrums on a periodic chain with the length $N=3N_u$. The instantaneous energy spectrums on open chains with different lengths: (b) $N=3N_u$; (c) $N=3N_u+1$; (d) $N=3N_u+2$. Here $N_u=20$, $\delta t_0=0.5$ and $W_0=1$.}\label{fig4}
\end{figure}

\begin{figure}[htbp]
\centering
\includegraphics[width=8.5cm]{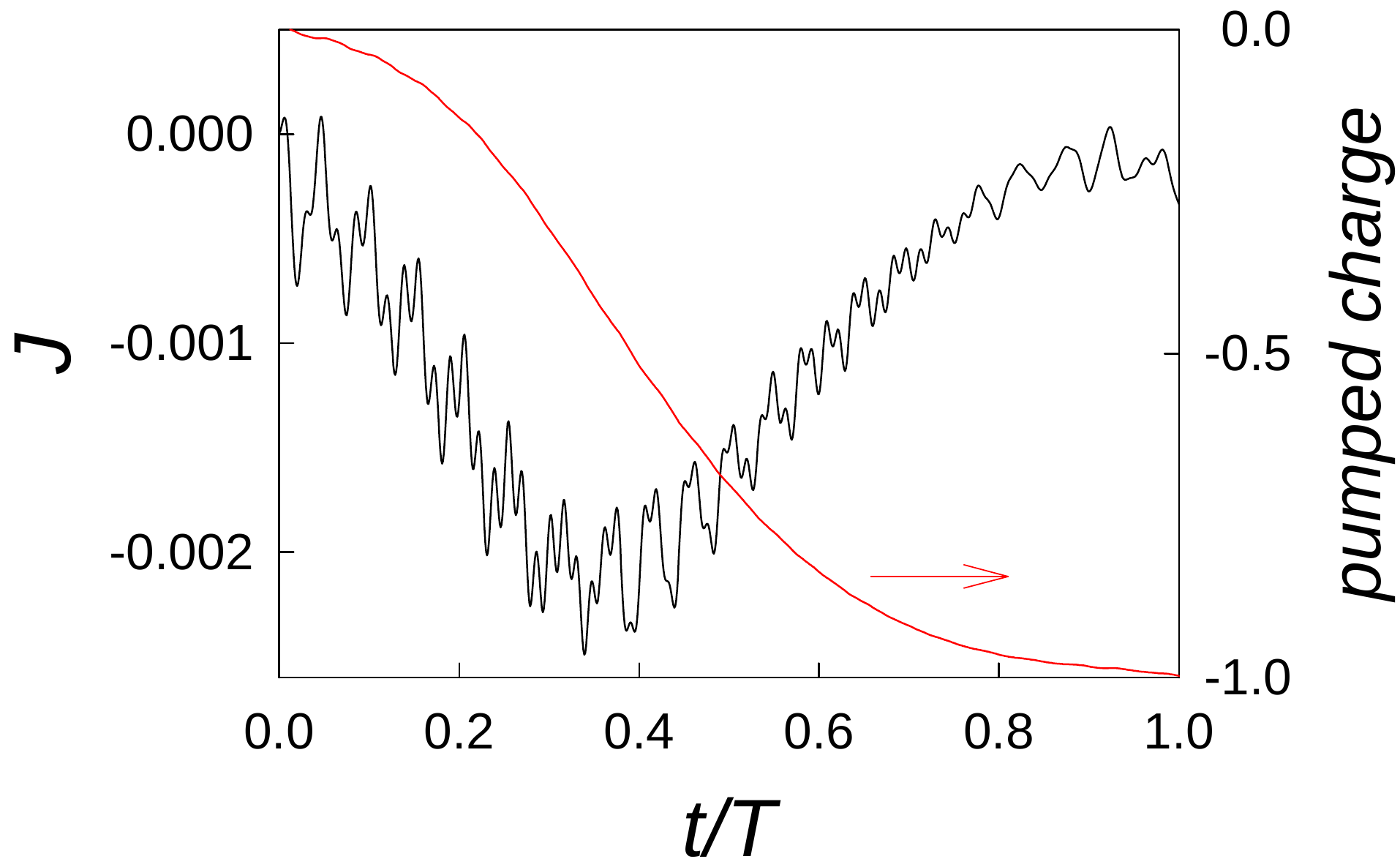}
\caption{(Color online) The instantaneous local current and the pumped charge on a periodic chain with the length $N=3N_u$ at $\frac{1}{3}$ filling. The parameters are the same as those in Fig.\ref{fig4}.}\label{fig5}
\end{figure}

In a cyclic adiabatic evolution of a one-dimensional insulator: $H(k,t+T)=H(k,t)$, the charge pumped across the insulator is a integer, which can be defined as a topological invariant \cite{rev20,rev21,rev22}. The process can be created in model Eq.(\ref{eq1}) by letting $\delta t=\delta t_0 \cos \frac{2\pi t}{T}$ and adding a on-site term $h_{st}=h_0 \sin \frac{2\pi t}{T}$, which has been discussed in previous works. In the following we create such processes in the models described by Eq.(\ref{eq2}), Eq.(\ref{eq3}) and Eq.(\ref{eq4}) respectively.

To create such a process in Eq.(\ref{eq2}), we let $\delta t=\delta t_0 \cos \frac{2\pi t}{T}$ in the first configuration of trimerization and add a on-site term $V_{on}=diag\{W,0,-W\}$ with $W=W_0\sin \frac{2\pi t}{T}$ the strength of the potential. The including of $V_{on}$ is to assure the system gapped when $\delta t=0$ at $t=T/4,3T/4$.
The instantaneous energy spectrums are shown In Fig.\ref{fig4}. The system is gapped on periodic chain, as shown in Fig.\ref{fig4} (a). While on open chains there appear gapless state traversing the gap, which are different for different lengths: $N=3N_u$ in Fig.\ref{fig4}(b), $N=3N_u+1$ in Fig.\ref{fig4}(c), $N=3N_u+2$ in Fig.\ref{fig4}(d). With the help of these gapless states, the integral charge is pumped across the insulator in a cyclic adiabatic evolution.

To show the details of the process, we directly calculate the current and the number of particles which have flown through the chain between the starting time and time $t$  is expressed as the integral of the current ${\cal J}_j$ at site $j$: $\Delta n(t)=\int_0^t dt' {\cal J}_j(t')$, where ${\cal J}_j=\langle \psi(t)|\hat{{\cal J}}_j|\psi(t)\rangle$ being the expectation value of the current operator $\hat{{\cal J}}_j$ \cite{note1}. The operator $\hat{{\cal J}}_j$ can be obtained vis the continuity equation \cite{note2}. For the process in Eq.(\ref{eq2}), $\hat{{\cal J}}_j=-i t_j (c_j^{\dagger}c_{j-1}-c_{j-1}^{\dagger}c_{j})$ with $t_j$ the total hopping amplitude between sites $j$ and $j-1$. In Fig.\ref{fig5} the instantaneous local current and the total pumped charge on a system with PBC at $\frac{1}{3}$ filling is shown and the result shows that one charge is pumped in a cycle. For the case of $\frac{2}{3}$ filling, the result is similar.

For a periodic system, varying the momentum $k$ and the time $t$, we get a manifold of Hamiltonian $H_2(k,t)$. If in a cycle the system remains gapped, the Chern number can be defined as \cite{rev23,rev24}:
\begin{eqnarray}\label{eq5}
C=\dfrac{1}{2\pi i}\int_{0}^{2\pi}dk \int_{0}^{T}dt F_{12}(k,t).
\end{eqnarray}
Here the field strength $F_{12}(k,t)=\partial_{k} A_2(k,t)-\partial_{t} A_1(k,t)$ with the Berry connection $A_{1(2)}(k,t)=\langle n(k,t) |\partial _{k (t)}|n(k,t) \rangle$ and $|n(k,t)\rangle$ the normalized wave function.  Using an numerical approach suggested in Ref.(26), the Chern number of the system in Fig.\ref{fig4}(a) is calculated. We find that the three bands have the Chern numbers $-1,0,1$ (from low energy to high). Thus the Chern number are $-1$ both at $\frac{1}{3}$ and $\frac{2}{3}$ fillings. The non-zero Chern number is consistent with the appearance of gapless states and the pumped integer charge, which manifest the nontrivial topology in the system.

\begin{figure}[htbp]
\centering
\includegraphics[width=8.5cm]{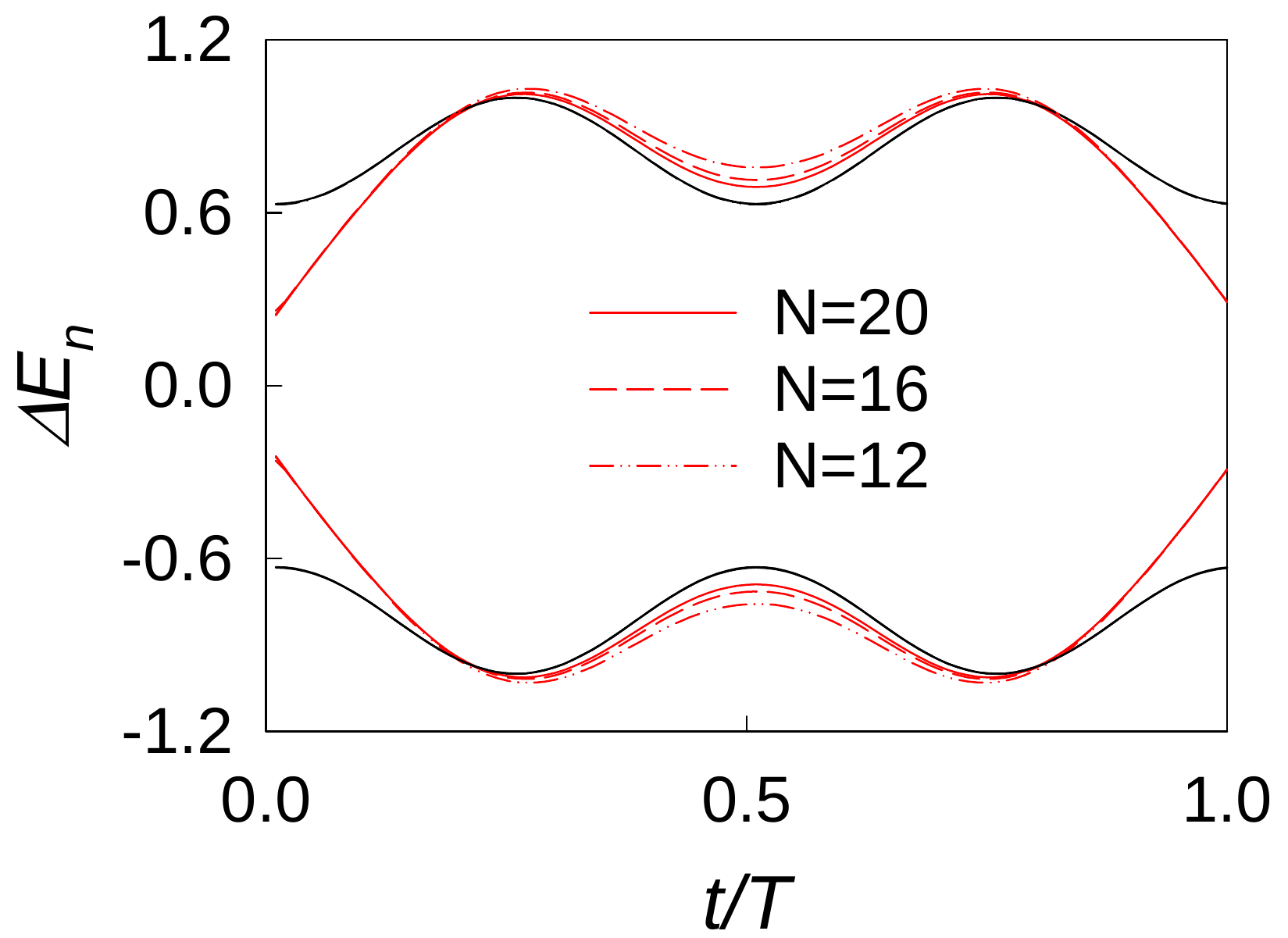}
\caption{(Color online) The energy spectrum of the quasiparticle added to a half-filled system with PBC (black) and OBC (red). The gap of the red line at $t/T=0,1$ is due to the finite-size effect. The parameters $V_1=W_0=1$.}\label{fig6}
\end{figure}

\begin{figure}[htbp]
\centering
\includegraphics[width=8.5cm]{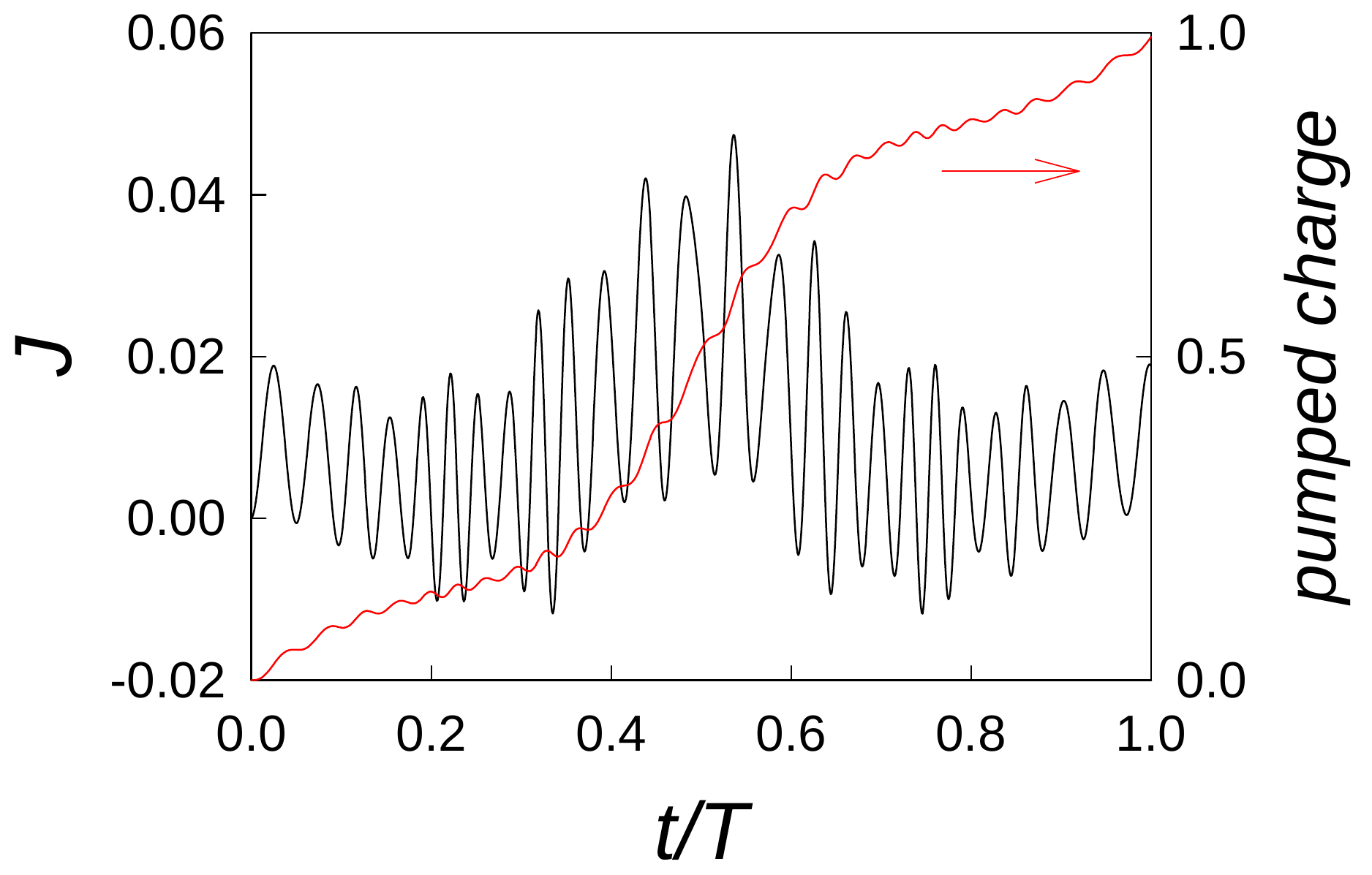}
\caption{(Color online) The instantaneous local current and the pumped charge on a periodic chain with the length $N=16$ at half filling for the pumping process in Eq.(\ref{eq3}). The parameters are the same as those in Fig.\ref{fig5}.}\label{fig7}
\end{figure}

To create such a process in Eq.(\ref{eq3}), we let $V_{ij}=(-1)^i V_1 \cos \frac{2\pi t}{T}$ and add an alternating on-site term $V_{on}=(-1)^{(i-1)} W_0 \sin \frac{2\pi t}{T}$. The energy spectrum of the quasiparticle added to a half-filled system
is shown in Fig.\ref{fig6}. When the boundary condition is periodic, the system is gapped. When the boundary condition is changed to open ones, there appear gapless states in the gap due to the nontrivial topology of the system. The instantaneous local current and the total pumped charge on a system with PBC at half filling are shown in Fig. \ref{fig7}, and one charge is pumped in a cycle. The Chern number of the interacting system can be calculated using the twisted boundary phase $\theta$, i.e., replacing the momentum $k$ with $\theta$ in Eq.(\ref{eq5}). For the process shown in Fig.\ref{fig6}, the Chern number is $1$.

\begin{figure}[htbp]
\centering
\includegraphics[width=8.5cm]{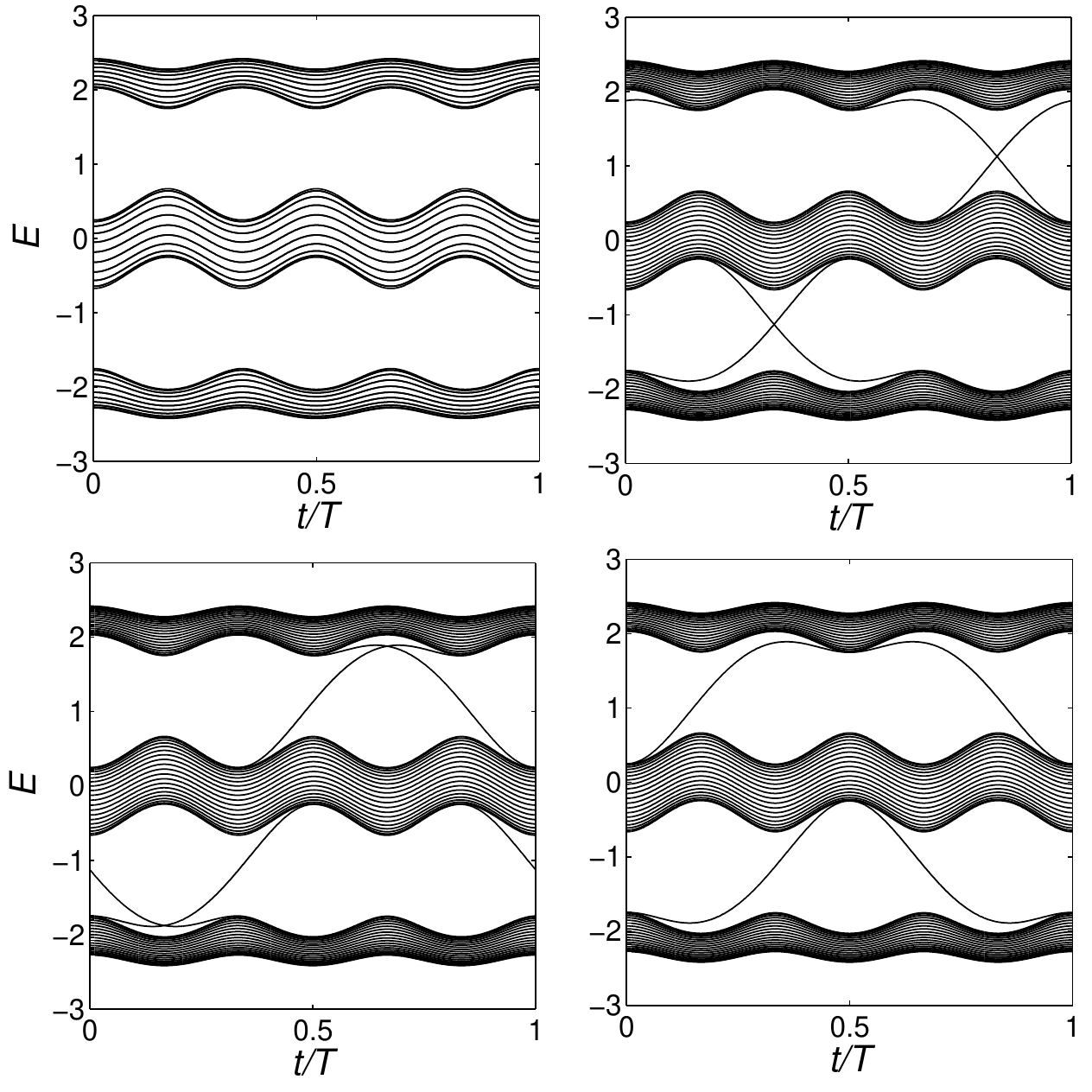}
\caption{(a) The instantaneous energy spectrums on a periodic chain with the length $N=3N_u$. The instantaneous energy spectrums on open chains with different lengths: (b) $N=3N_u$; (c) $N=3N_u+1$; (d) $N=3N_u+2$. Here the calculations are for the pumping process in Eq.(\ref{eq4}) and the parameters: $N_u=20$, $V=1.5$.}\label{fig8}
\end{figure}

To create such a process in Eq.(\ref{eq4}), we let the optical superlattice potential $V_i=V\cos(\frac{2\pi}{3} i+t)$ changing with the time. During a cycle the system remains gapped at $\frac{1}{3}$ and $\frac{2}{3}$ filling on a periodic chain [Fig.\ref{fig7} (a)], while there appear gapless states on a open chain. The instantaneous energy spectrums of open chains with different lengths are calculated: $N=3N_u$ in Fig.\ref{fig7}(b), $N=3N_u+1$ in Fig.\ref{fig7}(c), $N=3N_u+2$ in Fig.\ref{fig7}(d), where gapless states appear between the bands. By calculating the instantaneous local current and the total pumped charge on a system with PBC at $\frac{1}{3}$, we show in Fig. \ref{fig8} that one charge is pumped in a cycle. The Chern numbers of the three bands are $-1,2,-1$ respectively. So for this pumping process, the Chern number are $-1$ at $\frac{1}{3}$ filling, while $1$ at $\frac{2}{3}$ filling.

\begin{figure}[htbp]
\centering
\includegraphics[width=8.5cm]{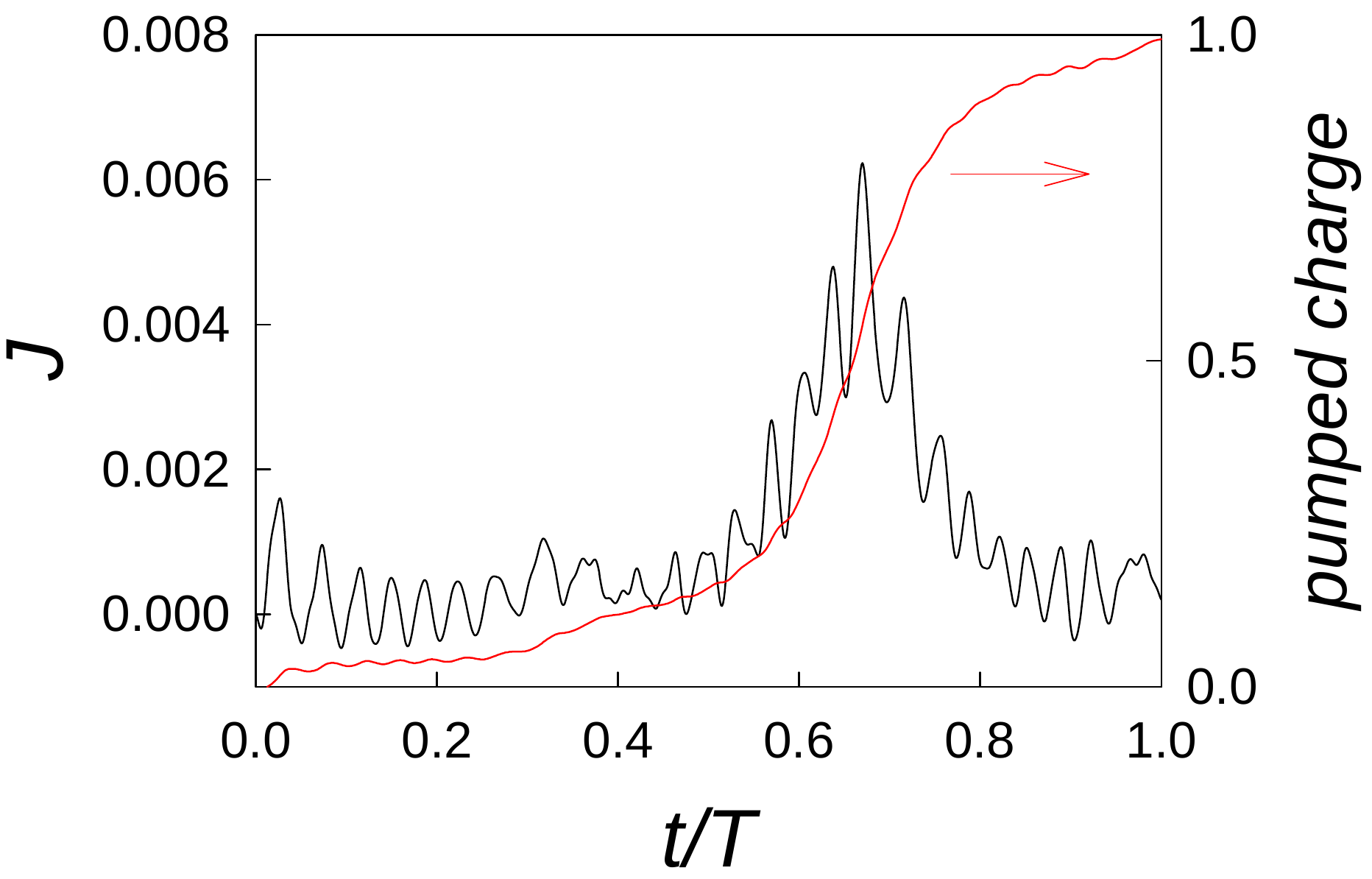}
\caption{(Color online) The instantaneous local current and the pumped charge on a periodic chain with the length $N=3N_u$ at $\frac{1}{3}$ filling for the pumping process in Eq.(\ref{eq4}). The parameters are the same as those in Fig.\ref{fig7}.}\label{fig9}
\end{figure}

\section{The conclusions}
We study 1D topological models with dimerization and trimerization and show that these models can be generated using interaction or optical superlattice. Firstly the dimerized and trimerized models are revisited. The study of the topological properties in these models are based on the appearance of edge states on open chains. Next we drop $\delta t$ terms in both models and dynamically generate $\delta t$ from interaction or optical superlattice. The mechanism is analyzed, which shows that the underlying physics is closely related to the dimerized and trimerized models. Then for each 1D topological model, a quantum pumping process is constructed. We calculate the instantaneous energy spectrum and local current, with which the quantum pumping process is explicitly demonstrated. We shows that the pumping is assisted by the gapless states connecting the bands and one charge is pumped during a cycle. In the process a nonzero Chern number can be defined. Our study systematically shows the connection of 1D topological models and quantum pumping, and is useful for the experimental studies on topological phases in optical lattices and photonic quasicrystals.

\section{Acknowledgements}
The author would like thank Shiping Feng and Shun-Qing Shen for helpful discussions. The work is supported by NSFC under Grant Nos. 11274032, 11104189 and Program for NCET.


\begin{thebibliography}{10}
\bibitem{note2}
The time derivative of the particle number on site $j$: $\frac{\partial c_j^{\dagger}c_j}{\partial t}=\frac{1}{i\hbar}[c_j^{\dagger}c_j,H]=\hat{{\cal J}}_j-\hat{{\cal J}}_{j+1}$, which determines the formula of the current operator.
\end{thebibliography}
\end{document}